\documentclass[prd,twocolumn,preprintnumbers,amsmath,amssymb]{revtex4}

\usepackage{citesort,latexsym,ifthen,graphics,color,hyperref}

\newcommand{\pder}[2]{\frac{\partial #1}{\partial #2}}
\newcommand{\fder}[2]{\frac{\delta #1}{\delta #2}}
\newcommand{\vect}[1]{\mathbf{#1}}
\newcommand{\Int}{\int\!\!}

\newcommand{\duv}{\Delta_{\mathrm{UV}}}
\newcommand{\dir}{\Delta_{\mathrm{IR}}}
\newcommand{\invdir}{\Delta_{\mathrm{IR}}^{-1}}
\newcommand{\cuv}{C_{\mathrm{UV}}}

\newlength{\IntHeight}
\newcommand{\Strut}[1]{\rule{0mm}{#1}}

\def\eq#1{(\ref{eq:#1})}

\def\Eqn#1{Equation~(\ref{eq:#1})}

\newcommand{\NuclPhys}[4]{{Nucl.\ Phys.\ }\textbf{#1 #2} (#3) #4}
\newcommand{\PhysRev}[4]{{Phys.\ Rev.\ }\textbf{#1 #2} (#3) #4}
\newcommand{\IntJModPhys}[4]{{Int.\ J.\ Mod.\ Phys.\ }\textbf{#1 #2} (#3) #4}

\newcommand{\PhysRept}[3]{{Phys.\ Rept.\ }\textbf{#1} (#2) #3}

\newcommand{\PhysLett}[4]{{Phys.\ Lett.\ }\textbf{#1 #2} (#3) #4}

\newcommand{\ProgTheorPhysS}[3]{{Prog.\ Theor.\ Phys.\ Suppl.\ }\textbf{#1} (#2) #3}

\newcommand{\arxiv}[1]{#1}
\newcommand{\hepth}[1]{hep-th/#1}
\newcommand{\hepph}[1]{hep-ph/#1}

\newcommand{\condmat}[1]{cond-mat/#1}
\newcommand{\astroph}[1]{astro-ph/#1}

\newcommand{\jphysa}[3]{J.\ Phys.\ {\bf A} #1 (#2) #3}

\newcommand{\arXiv}[2]{arXiv:{#1} [#2]}

\begin{document}

\preprint{DIAS-STP-07-18}

\title{A Comment on the Path Integral Approach to Cosmological Perturbation Theory
}

\author{Oliver J.~Rosten}
\email{orosten@stp.dias.ie}
\affiliation{Dublin Institute for Advanced Studies, 10 Burlington Road, Dublin 4, Ireland}

\begin{abstract}
	It is pointed out that the exact renormalization group approach to cosmological perturbation theory, proposed in Matarrese and Pietroni, JCAP {\bf 0706} (2007) 026,
\arxiv{\astroph{0703563}} and \arxiv{\astroph{0702653}}, constitutes a misnomer. Rather, 
having instructively cast this classical problem into path integral form,
the evolution equation then derived comes about as a special case of considering how the generating functional responds to variations of the primordial power spectrum.
\end{abstract}

\maketitle

Cosmological perturbation theory received a new lease of life when Crocce and Scoccimarro 
demonstrated, in a diagrammatic tour-de-force, how the perturbation series could be reorganized~\cite{Crocce1,Crocce2}. In the context of the evolution cold dark matter,
the central object in their approach is the full non-linear propagator, which measures the ensemble averaged response of the final density and velocity perturbations to variations in the initial conditions.\footnote{It is the process of ensemble averaging which produces loop diagrams, characteristic of quantum field theory, in this classical problem.} Not only does the reorganization make the encapsulation of the physics by the perturbation series more intuitive, but it also results in a much better behaved expansion. Crucially, by making some well motivated approximations, Crocce and Scoccimarro were able to resum the full propagator in the short distance limit and then interpolate between this regime and the long distance one to obtain an approximate form for the full propagator, at all scales.

Inspired by this, Matarrese and Pietroni~\cite{M+P,M+PLett} cast the (classical) problem into path integral form. The end point of their instructive demonstration is a generating functional which encodes the complete physics of the underlying equations, which themselves follow from applying the single stream approximation to the Vlasov equation. This approximation results in two coupled equations (the continuity and Euler equations) for the density perturbation, $\delta$, and peculiar velocity divergence, $\theta$.
Following Crocce and Scoccimarro, but using the notation of Matarrese and Pietroni, this pair of equations can be combined into a single equation which, when written in Fourier space (with repeated momentum arguments integrated over), takes the form:
\begin{eqnarray}
\nonumber
	\lefteqn{(\delta_{ab} \partial_\eta + \Omega_{ab}) \varphi_b(\vect{k}, \eta) =}\\
	&&
\label{eq:both}
	\qquad \qquad e^{\eta} \gamma_{abc}(\vect{k},-\vect{p},-\vect{q})\varphi_b(\vect{p},\eta) \varphi_c(\vect{q},\eta).
\end{eqnarray}
For the fine details, the reader is referred to~\cite{M+P}; for our purposes we note the following.
The doublet $\varphi_a$ is proportional to $\delta$ when $a=1$ and $\theta$ when $a=2$. 
$\gamma_{abc}$ is a vertex function, which couples together different modes. In an Einstein-de Sitter
cosmology, $\eta \sim \ln a$ where $a$ is the cosmological scale factor and 
$\Omega_{ab}$ is a constant matrix. (For generalizations to other cosmologies, see~\cite{Crocce2}.)

Now, assuming Gaussian initial conditions, Matarrese and Pietroni demonstrated that the physics of~\eq{both} is encoded by the following generating functional (all momentum dependence and some of the $\eta$ dependence is suppressed, for brevity):
\begin{eqnarray}
\label{eq:GF}
\lefteqn{Z[J_a ,K_b; P^0] =\Int \mathcal{D}\varphi_a \mathcal{D}\chi_b \exp \left\{ \Strut{\IntHeight} \right.
}  
\\ &&
		\Int d\eta d\eta'  
		\left[
			-{\textstyle \frac{1}{2}} \chi_a P^{0}_{ab} \delta(\eta) \delta(\eta') 
			\chi_b + i \chi_a g_{ab}^{-1} \varphi_b 
		\right] \qquad
\nonumber
\\ &&
		\qquad
		\left.
		\Strut{\IntHeight}
		-i \Int d\eta
		\left[
		e^\eta \gamma_{abc} \chi_a \varphi_b \varphi_c - J_a \varphi_a - K_b\chi_b
		\right]
		\right\}. \qquad
\nonumber
\end{eqnarray}
$\chi_a$ is an auxiliary doublet field, which carries information about the initial conditions, which are specified through the primordial power spectrum, $P^0$. The linear propagator, $g_{ab}$ [which can be obtained by solving the linearized version of~\eq{both}], couples the initial conditions to the final state, as expected~\cite{Crocce1}. $J_a$ and $K_b$ are sources.

The strategy which Matarrese and Pietroni claimed to follow was to use~\eq{GF} to derive an Exact Renormalization Group (ERG) equation. Whilst there is nothing mathematically incorrect about what was done,  it nevertheless does not amount to an implementation of the ERG; rather, what they did was a special case of noticing that an evolution equation can be derived from~\eq{GF} by examining variations with respect to the primordial power spectrum. Despite the initial motivation for doing this being, perhaps, somewhat suspect, it should be noted that the approach as a whole is certainly useful; in particular, it allows the machinery of functional techniques to be applied to the problem at hand. In so doing, Matarrese and Pietroni were able to reproduce Crocce and Scoccimarro's large-$k$ resummation of the propagator in a much simpler way and also derive their own expression for the full propagator at all scales, using transparent approximations.

In Quantum Field Theory (QFT), the central idea of the ERG is to integrate out degrees of freedom in such a way that the partition function---and hence the physics derived from it---remains invariant. As a first step in this procedure, an overall momentum cutoff is applied to the theory (in Euclidean space), with the action at this scale being the bare action. Next, one considers integrating out (coarse-graining) degrees of freedom between the bare scale and a lower, `effective' scale, $\Lambda$. The action at the effective scale is called the Wilsonian effective action, $S_\Lambda$ (which we here take to denote just the effective interactions, and not the regularized kinetic term as well). By considering how the Wilsonian effective action must evolve as $\Lambda$ changes,  if the partition function is to stay the same, one can derive the ERG equation, which describes how the effective action changes under changes of the effective scale~\cite{Wil,W+H,Pol,TRM-ApproxSolns,TRM-Elements}. Working with the scalar field, $\varphi$, we partition the standard propagator, $\Delta(q)= 1/q^2$, into an ultraviolet (UV) regularized part, $\duv$, and an infrared (IR) regularized part, $\dir$, such that $\Delta = \duv + \dir$.\footnote{Given a rapidly decaying cutoff function, $\cuv(q^2/\Lambda^2)$, we would write $\duv(q,\Lambda)= \cuv(q^2/\Lambda^2)/q^2$.} Polchinski's form of the ERG reads~\cite{Pol}:
\begin{equation}
\label{eq:Pol}
\pder{S_\Lambda[\varphi]}{\Lambda} =
\frac{1}{2} \fder{S_\Lambda}{\varphi} \cdot \pder{\duv}{\Lambda} \cdot  \fder{S_\Lambda}{\varphi}
-
\frac{1}{2} \fder{}{\varphi} \cdot \pder{\duv}{\Lambda} \cdot  \fder{S_\Lambda}{\varphi}.
\end{equation}

By performing a Legendre transform of the Wilsonian effective action, the ERG equation can be  transformed into a flow equation for the (IR regulated) generator of 1PI diagrams, $\Gamma_\Lambda$, the `effective average action'~\cite{TRM-ApproxSolns,TRM-Elements,Wetterich}:
\begin{equation}
\label{eq:EAA}
	\pder{\Gamma_{\Lambda, \mathrm{int}}[\varphi^c]}{ \Lambda} = 
	\frac{1}{2}
	\mathrm{tr}\!
	\left[
		\pder{\invdir}{\Lambda} \cdot
		\left(
			\invdir +  \Gamma_{\Lambda,\mathrm{int}}^{(2)}  
		\right)^{-1}
	\right],
\end{equation}
where $\varphi^{c}$ is the classical field, the trace indicates a momentum integral, the superscript `$(2)$' indicates a double functional derivative with respect to $\varphi^{c}$, and `int' denotes the interaction part of $\Gamma_\Lambda$.

Let us contrast the ERG approach to that of Matarrese and Pietroni. Their first step is to introduce a high frequency cutoff in the primordial power spectrum, at a scale $\lambda$. We emphasise that this is a restriction on the boundary conditions and not on the final perturbations: the non-linear interactions can generate power at the `missing' scales. The next step simply amounts to observing that, since the generating
functional~\eq{GF} depends on the primordial power spectrum, it now depends on $\lambda$. By differentiating with respect to $\lambda$,  Matarrese and Pietroni obtain the following evolution equation (equation~(52) of~\cite{M+P}):
\begin{equation}
\label{eq:M+P-1PI}
	\partial_\lambda {\Gamma}_{\lambda,\mathrm{int}} = \frac{i}{2}
	\mathrm{tr}\!
	\left[
		\left(\partial_\lambda \vect{\Gamma}^{(2)}_{\lambda,\mathrm{free}} \right) \cdot
		\left(
			\vect{\Gamma}^{(2)}_{\lambda,\mathrm{free}} + \vect{\Gamma}^{(2)}_{\lambda,\mathrm{int}}
		\right)^{-1}
	\right],
\end{equation}
where `free' denotes the free part of the appropriate object.
The double derivatives are with respect to combinations of $\chi_a$ and $\phi_b$ and so form a matrix, and the trace now also includes an integral over $\eta$ and summations over the doublet indices. 
\Eqn{M+P-1PI} is clearly of a very similar form to~\eq{EAA}. However, it should be emphasised that there has been no coarse-graining of modes; indeed, it is not intuitively clear what such a procedure would amount to in this scenario. Furthermore, the non-locality of the three-point vertex anyway indicates that the ERG is the wrong language to be using, since a fundamental requirement 
of the ERG is that the Kadanoff blocking (coarse-graining) transformation only affects variables in a localized patch~\cite{Wil,ym1}. Finally, we note that, contrary to the ERG approach, where the introduction of $\Lambda$ is central, the introduction of $\lambda$ is actually not necessary: the results of~\cite{M+P,M+PLett} can be derived simply by considering general variations of $P^0$.%
\footnote{
In terms of classifying the approach of Matarrese and Pietroni, `RG-inspired' would seem to be the
appropriate terminology. Note that the RG has proven itself a powerful tool in classical problems,
see for example~\cite{Chen}.}

Consequently, none of the intuition behind the ERG nor, for example, the powerful derivative expansion which is so fruitfully applied within this framework (see~\cite{B+B} for a review) are appropriate to cosmological perturbation theory.\footnote{The derivative expansion relies on Taylor expanding the flow equation in 
external momenta, which is automatically spoilt by the nonlocal three-point vertex.} On the other hand, the generating functional~\eq{GF} seems the perfect device with which to efficiently understand the resummations of Crocce and Scoccimarro and, one might hope, provides an starting point for future study. For interesting possible directions, see~\cite{Future}. For a bona-fide application of the ERG in a cosmological context, see~\cite{JG}.
 
 \begin{acknowledgements}
 	It is a pleasure to thank Mart\'{\i}n Crocce for extremely helpful discussions and for comments on the manuscript, and Daniel Litim for some vigourous debating.
 \end{acknowledgements}

\end{document}